\documentclass[aps,prl,twocolumn,showpacs,groupedaddress,amsmath,amssymb]{revtex4-1}
\usepackage{graphicx}
\usepackage{dcolumn}
\usepackage{bm}
\usepackage{color}
\usepackage{hyperref}
\usepackage[caption=false]{subfig}
\usepackage{dsfont}

\hypersetup{colorlinks=true,urlcolor=blue,linkcolor=blue,citecolor=blue}

\newcommand{\bra}[1]{\left\langle#1\right\vert}
\newcommand{\ket}[1]{\left\vert#1\right\rangle}

\begin{document}

\title{Quantum state engineering with twisted photons via adaptive shaping of the pump beam}
\author{E.\,V.\,Kovlakov}\email{ekovlakov@gmail.com} \author{S.\,S.\,Straupe} \author{S.\,P.\,Kulik}
\affiliation{Quantum Technologies Centre, M.V.Lomonosov Moscow State
	University, Moscow, Russia}
\affiliation{Faculty of Physics, M.V.Lomonosov Moscow State
University, Moscow, Russia}

\date{\today}

\begin{abstract}
	High-dimensional entanglement is a valuable resource for quantum communication, and photon pairs entangled in orbital angular momentum are commonly used for encoding high-dimensional quantum states. However, methods for preparation of maximally entangled states of arbitrary dimensionality are still lacking, and currently used approaches essentially rely on filtering and entanglement concentration. Here we experimentally realize a method for generation of high-dimensional maximally entangled OAM states of photon pairs which does not require any of these procedures. Moreover, the prepared state is restricted to the subspace of the specified dimensionality, thus requiring minimal postselection. 
\end{abstract}


\maketitle
The use of high-dimensional entangled systems denoted as qudits in quantum communications offers a number of advantages over the well-studied qubit systems such as higher information capacity \cite{bechmann2000quantum, walborn2006quantum}, enhanced robustness against eavesdropping in quantum key distribution (QKD) protocols \cite{cerf2002security} and stronger violation of generalized Bell's inequalities \cite{kaszlikowski2000violations} with possible applications in device independent QKD \cite{BarrettPRL2005,VaziraniPRL2014} and randomness generation \cite{AcinNature2016}. 
To date, orbital angular momentum (OAM) of light produced by spontaneous parametric down-conversion (SPDC) has become the workhorse for two-qudit states generation \cite{erhard2017twisted}. Twisted photons have enabled qutrits encoding technique \cite{vaziri2002experimental}, which has been successfully used in a QKD protocol \cite{groblacher2006experimental}. OAM of photons has recently been successfully used for three-dimensional GHZ state generation \cite{erhard2017experimental}.

Ideally, in a nearly collinear phase-matching geometry, the angular momentum of photons in the down-conversion process is conserved \cite{mair2001entanglement}. Therefore, if a pump photon has zero OAM value, the produced two-photon state is anti-correlated in OAM:
\begin{equation}
\label{Psi}
\ket{\Psi} = \sum_{l=-\infty}^{+\infty} c_{l} \ket{l,{-}l}.
\end{equation}
Here $|c_{l}|^{2}$ determine the probabilities of finding a signal photon in the eigenstate $\ket{l}$ carrying $l \hbar$ units of OAM and an idler photon in the state $\ket{-l}$ carrying $-l \hbar$ units of OAM. The width of $|c_{l}|^{2}$ distribution is called the spiral bandwidth and depends on the crystal length and the pump beam waist \cite{torres2003quantum}. Since the amplitudes $c_{l}$ are in general non-equal and decrease with increasing $l$, the OAM-state (\ref{Psi}) is not maximally entangled. Therefore, the generated state requires a procedure of entanglement concentration \cite{bennett1996concentrating} to equalize these weights \cite{vaziri2003concentration, dada2011experimental}. This method (also referred to as ``Procrustean filtering'') implies the extraction of maximally entangled states out of nonmaximally entangled ones using spatial filtering, which inevitably leads to loss. In this Letter we present an experimental realization of maximally entangled two-qudit state generation without the need for such a filtering procedure. 

Previous experimental results have shown that the spatial mode spectrum of SPDC may be radically modified with the use of spatially shaped pump. In particular, lossless generation of spatial Bell-states using low-order Hermite-Gaussian pump beams \cite{romero2012orbital,kovlakov2017spatial} was reported. Here we go further and engineer high-dimensional entangled states with a much more complex pump beam transformation. The method we use is inspired by the theoretical work of Torres et al. \cite{torres2003preparation} who showed that the topological information imprinted in the pump light can be translated into the amplitudes of the generated entangled quantum states in a controlled way.
 
Following the idea of our previous experiment \cite{kovlakov2017spatial}, we first minimize the spiral bandwidth by the optimal pump beam focusing to concentrate the flux of the down-converted photons in the low-order modes subspace. Then we reconfigure the OAM spectrum of the SPDC radiation by converting an initially Gaussian pump beam into a superposition of Laguerre-Gaussian modes $\mathrm{LG}_{p}^{l}$ of width $w$: 
\begin{equation}
\label{Pump}
\mathcal{E}_{p}(\rho, \phi; w) = \sum_{l} \alpha_{l} \mathrm{LG}_{0}^{l}(\rho, \phi; w),
\end{equation}
where $\mathcal{E}_{p}(\rho, \phi; w)$ function describes the electric field of the transformed beam in cylindrical coordinates $\{ \rho,\phi \}$ and $\alpha_{l}$ are complex-valued coefficients. The index $l$ is associated with an azimuthal phase term $\exp{(il\phi)}$ of the LG beam and the radial index $p$ is taken to be zero. The careful adjustment of $\alpha_{l}$  allows us to control both the weights and the phases of maximally entangled high-dimensional states. Thus, the method presented here is a valuable alternative to the filtering approaches mentioned above and to the recently proposed technique of qudits generation based on entanglement by path identity \cite{krenn2017entanglement}. 

\begin{figure}
	\center{\includegraphics[width=0.95\linewidth]{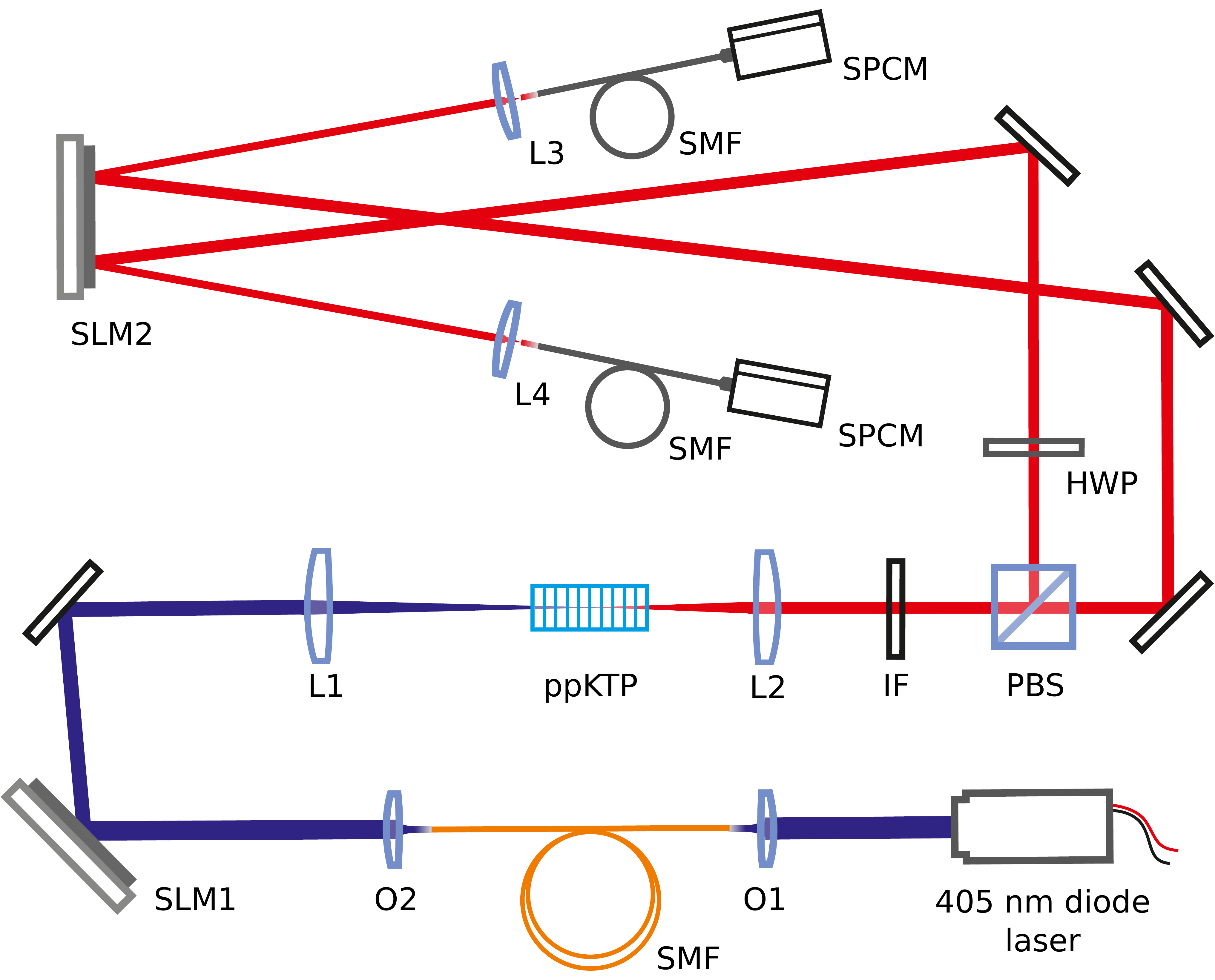}}
	\caption{Experimental setup: L1 = 200 mm, L2 = 100 mm, L3 = L4 = 11 mm; O1 and O2 -- 20x and 10x microscopic objectives, respectively; IF -- 810$\pm$5 nm interference filter; SPCM -- single-photon counting modules}\label{fig:Setup}
\end{figure}

\textit{Experiment.}---We use a 15-mm-thick periodically poled KTP crystal designed for a collinear frequency degenerate type-II phase matching as a source of entangled photon pairs. The output beam of grating-stabilized 405 nm diode laser is spatially filtered by a single mode fiber and then shaped by the first spatial light modulator SLM1 (Cambridge Correlators). The resulting field in the first diffraction order of the SLM1 is focused on the crystal via a lens L1 (see Fig.~\ref{fig:Setup} for the details). Since the signal and idler photons have orthogonal polarizations, they are separated by the polarizing beam splitter (PBS). We use a well-known scheme for projective measurements in the basis of LG modes \cite{agnew2011tomography} by focusing the signal and idler beams on the corresponding halves of an SLM2 (Holoeye Pluto) followed by single mode fibers and photon counting modules. A half wave plate (HWP) is inserted to optimize the polarization of photons reflected by the PBS for the second SLM. 
The holographic masks for the LG modes generation and detection are calculated according to the method described in \footnote[2]{See Supplemental Material at [URL will be inserted by publisher], which includes Ref. \cite{bolduc2013exact}, \cite{zhang2014radial}, \cite{KwiatPRA2001}, for a detailed discussion of holograms calculation, optimization algorithm, Bell inequalities for qutrits, maximally entangled qubits phase control, and full state tomography of entangled qutrits state \label{supplementary}}. 
\begin{figure}
	\center{\includegraphics[width=1.\linewidth]{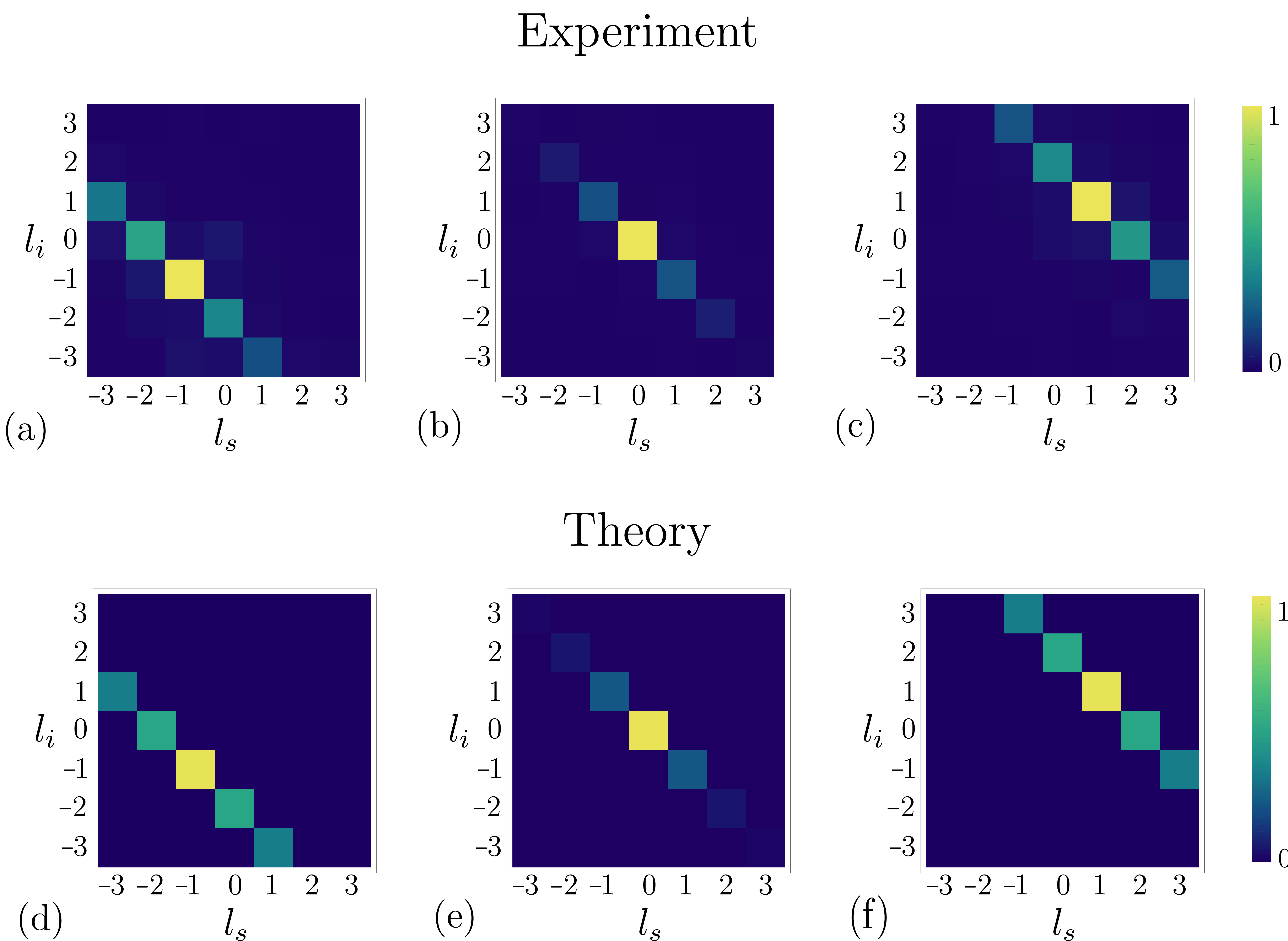}}
	\caption{Normalized (divided by the maximal value) coincidence count rates as a function of detection modes azimuthal numbers $l_{s}$ and $l_{i}$: experiment (top row) and theory for the azimuthal Schmidt number $K_{az} = 2$ (bottom row). The cases of the pump beam with $l=-2$ (left column), $l=0$ (center column) and $l=2$ (right column) are shown.}\label{fig:spectra}
\end{figure}

In optimizing the regime of down-conversion for a minimal spiral bandwidth, we follow the formalism of Schmidt modes developed in \cite{law2004analysis}. According to this concept, the pump waist is chosen such that the crystal length $L$ is approximately twice that of the Gaussian pump beam Rayleigh range. Thus we focus the pump beam to a waist size $w = \sqrt{L/k_{p}} \approx 25~\mu$m, where $k_{p}$ denotes the wave vector of the pump. For the optimal detection of the down-converted modes we use the detection beam waist $\sigma \approx 33~\mu$m, which is close to the theoretically optimal $\sigma = \sqrt{2} w$ for the single-Schmidt mode regime \cite{kovlakov2017spatial}. As a result, the experimentally measured azimuthal correlations between the idler and the signal channels reveal very low contribution of the down-converted photons with $|l|>1$ (see Fig.~\ref{fig:spectra}(b)). The number of azimuthal spatial modes can be estimated as $K_{az} = 1/\sum_{l} \lambda_{l}^{2}$, where the eigenvalues of the Schmidt decomposition $\lambda_{l}$ are equal to $|c_{l}|^{2}$ probabilities from (\ref{Psi})~\cite{miatto2012spatial}. Hence, we estimate the value of $K_{az} = 2.0 \pm 0.1$ from the diagonal distribution $l_{s} = - l_{i}$ of measured coincidence count rates. Such a small azimuthal Schmidt number allows us to decrease the number of unused photons in high-order modes during the further spiral spectrum reconfiguration.
\begin{figure*}
	\center{\includegraphics[width=0.95\linewidth]{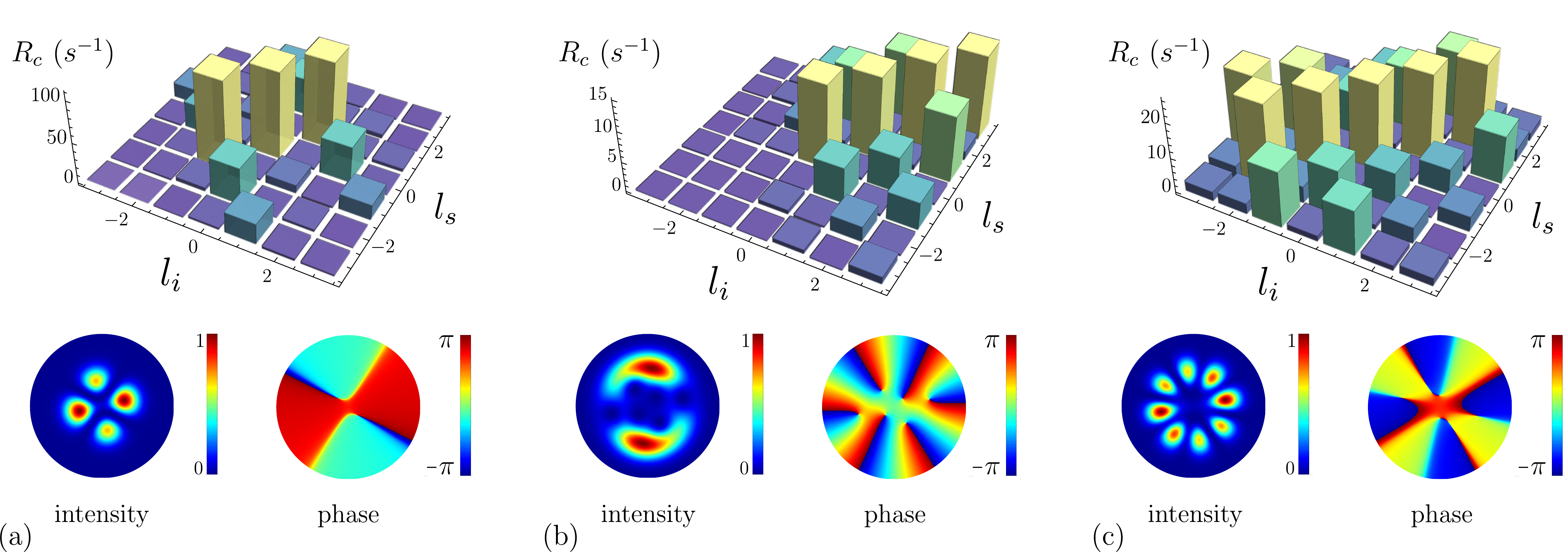}}
	\caption{Spiral spectra of maximally entangled (a) qutrits, (b) ququarts and (c) ququints. Associated intensity and phase profiles of the pump beams calculated from the experimentally obtained coefficients $\alpha_{l}$ in (\ref{Pump}) are shown in the bottom row. The corresponding non-zero $\alpha_{l}$ are (a) $\alpha_{-2}=0.76-0.11i,~\alpha_{0}=-0.12+0.15i,~ \alpha_{2}=0.30-0.53i$, (b) $\alpha_{0}=0.09-0.02i,~\alpha_{2}=-0.02-0.19i,~ \alpha_{4}=0.57-0.01i,~\alpha_{6}=0.77-0.21i$, (c) $\alpha_{-4}=-0.25-0.73i,~\alpha_{-2}=0.19-0.10i,~ \alpha_{0}=-0.07+0.11i,~\alpha_{2}=0.14-0.14i,~\alpha_{4}=-0.54+0.09i$.}\label{fig:Qudits}
\end{figure*}

\textit{Qudit state engineering.}---Due to the conservation of OAM, switching between an initially Gaussian pump beam and LG modes with $l=2$ or $l=-2$ leads to a shift of the down-converted modes distribution from the leading diagonal $l_{s} = - l_{i}$ to the upper-diagonal $l_{s} = 2 - l_{i}$ or sub-diagonal $l_{s} = -2 - l_{i}$, respectively (see Fig.~\ref{fig:spectra}(a,c)). Moreover, this distribution becomes wider, indicating that entanglement between spatial modes also increases with the increasing absolute value of pump OAM, being in a good agreement with our numerical calculations for the biphoton amplitude in the Gaussian approximation \cite{law2004analysis} shown in Fig.~\ref{fig:spectra}(d,e,f). As one can see from the presented spiral spectra, the use of the pump beam in a superposition of three even low-order LG modes may provide the cross-correlation histogram with three equal anti-diagonal elements, which corresponds to the generation of two maximally entangled qutrits with some phases $\theta_{1}$ and $\theta_{2}$:
\begin{equation}
\label{Qutrit}
\ket{\Psi^{(3)}} =\dfrac{\exp{(i \theta_{1})} \ket{\scalebox{0.8}[1.0]{-}1,\scalebox{0.8}[1.0]{-}1} + \ket{0,0} + \exp{(i \theta_{2})} \ket{1,1}}{\sqrt{3}} ,
\end{equation}
in the subspace $S_{3}=\{ \ket{\scalebox{0.8}[1.0]{-}1,\scalebox{0.75}[1.0]{-}1}, \ket{0,0}, \ket{1,1} \}$. 

To equalize the coefficients in the generated superposition precisely and to take into account the experimental errors associated with non-perfect overlap between the pump modes and the detection modes we further optimize the values of $\alpha_{l}$ coefficients with an adaptive procedure.

For this optimisation we use a simultaneous perturbation stochastic approximation (SPSA) algorithm introduced in \cite{spall1998implementation}. This algorithm requires only two cost function measurements at each iteration of an optimization process, regardless of the problem dimensionality. It means that we use only two proposal vectors of modal weights $\alpha = \{\alpha_{-2},\alpha_{0},\alpha_{2}\}$ and experimental estimates of cost function to provide the direction to the optimal pump beam configuration. As a cost function $f(\alpha)$, we choose a variance of three measured probabilities $|\bra{l_{s},l_{i}} \Psi^{(3)}(\alpha) \rangle |^{2} $ for $l_{s} = l_{i} = -1,0,1$, where the state vector of the generated state $\ket{\Psi^{(3)}(\alpha)}=a_{1}(\alpha)\ket{\scalebox{0.8}[1.0]{-}1,\scalebox{0.8}[1.0]{-}1} + a_{2}(\alpha) \ket{0,0} + a_{3}(\alpha) \ket{1,1}$ depends on the vector $\alpha$. In other words, we seek to minimize the difference between the absolute values of the measured amplitudes $a_{i}(\alpha)$ and equal weights $1/\sqrt{3}$ to produce the maximally entangled qutrits. The resulted OAM spectrum of a maximally entangled qutrits followed by the corresponding intensity and phase profiles of the pump are shown in Fig.~\ref{fig:Qudits}(a). The detailed description of the algorithm behaviour is given in the supplementary material \footnote[2]{}.

We have repeated the same procedure to produce maximally entangled ququarts by pumping the crystal with a superposition of four LG beams with even and positive OAM values $l=0,2,4$ only. The resulted state is maximally entangled in the subspace $S_{4}=\{ \ket{0,0}, \ket{1,1}, \ket{2,2}, \ket{3,3} \}$. The corresponding beam represents a ``vortex pancake'' -- a Gaussian beam with phase vortices nested in it (see Fig.~\ref{fig:Qudits}(b)). In analogy with the previous case, we rewrite the cost function for the adaptive optimization $f(\alpha)$ as a variance of the four measured probabilities $|\bra{l_{s},l_{i}} \Psi^{(4)}(\alpha) \rangle |^{2} $ for $l_{s} = l_{i} = 0,1,2,3$ with $\alpha = \{\alpha_{0},\alpha_{2},\alpha_{4},,\alpha_{6}\}$.

Finally, we have prepared maximally entangled ququints in the subspace $S_{5}=\{ \ket{\scalebox{0.8}[1.0]{-}2,\scalebox{0.8}[1.0]{-}2},\ket{\scalebox{0.8}[1.0]{-}1,\scalebox{0.8}[1.0]{-}1}, \ket{0,0}, \ket{1,1}, \ket{2,2} \}$ using a superposition of five LG beams with even $l=-4,-2,0,2,4$ as a pump. The obtained experimental results are shown in Fig.~\ref{fig:Qudits}(c). Here we need to note, that since in the ququints case we use LG beams both positive and negative indexes, the maximal mode order of the generated pump beam is lower than in the case of ququarts. This leads to a more efficient conversion of the initially Gaussian pump beam to the LG modes superposition despite the higher dimensionality of the prepared state. In particular, the power of the radiation incident on the crystal after the corresponding phase masks is 1.5 mW, 0.7 mW and 1.1 mW for the cases of qutrits, ququarts and ququints, respectively. At the same time, from the histograms presented in Fig.~\ref{fig:Qudits} we can conclude that the coincidence rates $R_{c}$ for the ququints case are nearly twice as high as for the ququarts one. Of course one can use a subspace other than the proposed $S_{4}$, for example, the exclusion of any LG mode from the ``ququints'' pump seems to be a more preferable way to generate ququart states. 

\textit{Qutrits phase control.}---Despite the fact that the measured spectra demonstrate the equality of the qudit amplitudes, they provide no information about the phases. Moreover, the question arises whether the pump light control allows one to produce a qudit with arbitrary phases at all. It is natural to assume that the global rotation of the pump beam does not influence the amplitudes due to the azimuthal symmetry of LG beams. According to the analytical predictions for the vortex pancake case, this statement is true \cite{torres2003preparation}. At the same time, the relative phases of the components of maximally entangled states vary with the pump beam rotation angle $\theta$ deterministically. We have numerically verified the following modification of Eq.~(\ref{Qutrit}) with respect to $\theta$:
\begin{eqnarray}
\label{QutritPhases}
\ket{\Psi^{(3)}(\theta)} =\dfrac{1}{\sqrt{3}}[ \exp{(i \theta_{1} - i 2 \theta)} \ket{\scalebox{0.8}[1.0]{-}1,\scalebox{0.75}[1.0]{-}1} + \ket{0,0}\nonumber \\ 
 + \exp{(i \theta_{2} +i 2 \theta)} \ket{1,1}].
\end{eqnarray}
\begin{figure}
	\center{\includegraphics[width=0.99\linewidth]{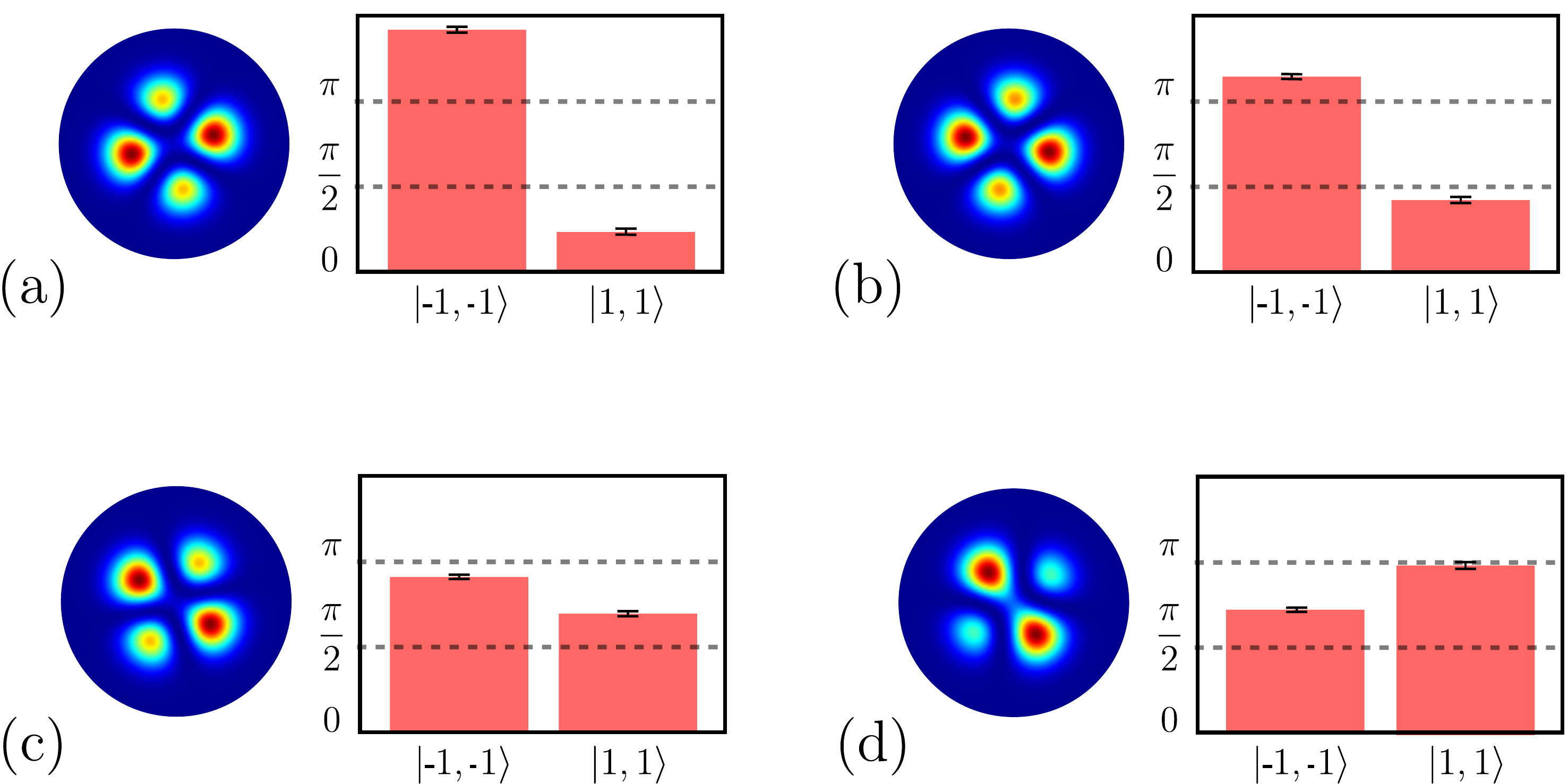}}
	\caption{Experimental phases of qutrits terms $\ket{-1,-1}$ and $\ket{1,1}$ for different rotation angles (a) $\theta = 0$, (b) $\theta = \pi/8$, (c) $\theta = \pi/4$ and (d) $\theta = 3\pi/8$ followed by corresponding intensity profiles of the pump beam. The phase of $\ket{0,0}$ term is taken to be zero as a reference. Phase errors are calculated from Monte Carlo simulations of Poissonian counting statistics.}\label{fig:phases}
\end{figure}

This prediction is in a good agreement with the experiment. In order to estimate the phases of the prepared qutrits, we perform a full quantum state reconstruction in a way described in \cite{agnew2011tomography} for a 9-dimensional OAM subspace, spanned by all possible pairwise tensor products of $\ket{l}$ vectors with $l=-1,0,1$ (see the supplementary material \footnote[2]{} for details). Since the reconstructed density matrix $\rho$ is mixed (as implicitly assumed by the chosen parametrization), we chose its eigenvector with the largest eigenvalue ($\approx 0.92$) as an estimate of the closest pure state, and compare it with (\ref{QutritPhases}). The phases of $\ket{\scalebox{0.8}[1.0]{-}1,\scalebox{0.8}[1.0]{-}1}$ and $\ket{1,1}$ terms obtained experimentally for varying $\theta$ are presented in Fig.~\ref{fig:phases}. After the rotation of the pump beam, the amplitudes of the qutrit components become slightly unequal. We launch our adaptive algorithm after each rotation by an angle $\pi/8$ to equalize the amplitudes again. As a result, the pump beam intensity redistributes across the beam with its rotation, however, these changes are barely visible. Interestingly, in the case of qubits, the same approach allows us to produce the states with a completely arbitrary phase \footnote[2]{}.

\textit{Entanglement verification.} --- To further demonstrate the entanglement of the generated qutrits, we have made use of the Collins-Gisin-Linden-Massar-Popescu (CGLMP) inequalities, which are the Bell inequalities generalized for the $d$-dimensional case \cite{collins2002bell}. It was shown, that the Bell parameter $I_{d=3}$ has to be less than 2 for any local realistic theory and is approximately equal to $2.87$ for the case of maximally entangled qutrits. We experimentally measure the value of $I_{3}=2.56 \pm 0.06$, which is well above the classical limit, but lower than the theoretical upper bound. We attribute the reduction of $I_{3}$ mainly to the modest value of purity for the experimentally generated state $\mathrm{Tr}\rho^{2} = 0.85 \pm 0.02$. Relatively low values of purity seem to be caused by the imperfections of our mode detection technique and are ubiquitous for such realizations of projective measurements in the spatial modes basis. In addition, the grating defects in the periodically poled crystal, which are known to affect negatively the single-mode coupling efficiency \cite{fedrizzi2007wavelength}, may also reduce the purity of the prepared state.

It is well known, that the CGLMP inequalities for the dimensionality $d>2$ are maximally violated by non-maximally entangled states. In particular, for qutrits the maximal violation is obtained for the states of the form $\ket{\Psi}=1/\sqrt{2+\gamma^2}\left(\ket{\scalebox{0.8}[1.0]{-}1,\scalebox{0.75}[1.0]{-}1} + \gamma\ket{0,0} + \ket{1,1}\right)$ \cite{acin2002quantum}. The maximal value of $I_3\approx 2.91$ corresponds to $\gamma\approx 0.79$. We prepared this state experimentally (see supplementary materials for full tomography of the corresponding state), however, we were not able to significantly improve the violation -- the experimental value of $I_3= 2.61 \pm 0.05$ is equal to that of maximally entangled states within the experimental uncertainty. This value is in agreement with theoretical predictions for the uncolored noise model $\rho = p \ket{\psi}\bra{\psi}+(1-p) \mathds{1} / d^2$, where the $I_3$ parameter scales as $p I_3$. The experimentally obtained $p = 0.91 \pm 0.02$ and $p = 0.88 \pm 0.02$ for the maximally and non-maximally entangled states, correspondingly, explain well the observed reduction of $I_3$, which was also reported before in other experiments \cite{bernhard2013shaping}.

\textit{Discussion.}---We have experimentally demonstrated a method for the generation of spatially entangled states of photons with variable dimensionality. In this Letter we mostly focused on generating maximally entangled states with equal amplitudes of the components in the superposition, however the method is completely general, and may be used to generate qudit states with arbitrary distribution of amplitudes. The level of control over generated states demonstrated here is sufficient, for example, to generate all mutually unbiased bases for a realization of a high-dimensional QKD protocol \cite{bechmann2000quantum,cerf2002security}. Moreover, the adaptive procedure used here to supplement the analytical heuristic may be utilized on its own to generate completely arbitrary spatial states of photon pairs, with the only limitation being the conservation laws in the SPDC process. For example, one may use full state tomography to estimate the fidelity of the prepared state with the desired one, and use it as a cost function for the optimization routine. We believe that this approach may become an interesting and fruitful research direction.   

This work was supported by the Russian Science Foundation project 16-12-00017. EVK acknowledges support form the BASIS foundation.

\textit{Note added.} After the completion of this manuscript we have become aware of the closely related work \cite{liu2018coherent}.

\bibliographystyle{apsrev4-1}
\bibliography{paper} 

\clearpage

\onecolumngrid
\section*{\large Supplementary Information}
\twocolumngrid

\section*{Holograms calculation}

Our method of holograms calculation is based on the algorithm presented in \cite{bolduc2013exact}: the phase profile imprinted on the hologram contains the phase distribution of the desired field and a blazed grating pattern modulated by the desired amplitude distribution. This method implies that the input field is a plane wave, so its direct application to a Gaussian beam with a finite waist causes some unwanted amplitude alteration. The reverse process of mode selection with a single-mode fiber also requires us to take into account the difference between the plane wave and the fundamental fiber mode \cite{zhang2014radial}. Thus to generate an LG mode $\mathrm{LG}_{0}^{l}(\rho, \phi, w) \propto (\rho / w)^{|l|}\mathrm{L}^{|l|}_{0}(2 \rho^{2}/w^{2}) \exp{(-\rho^{2}/w^{2})} \exp{(il\phi)}$ (where $\mathrm{L}^{|l|}_{0}$ is an associated generalized Laguerre polynomial) we use a modified expression for the field imprinted on the hologram, with the modified waist $\tilde{w}$ introduced to take into account the finite incident beam waist and to avoid amplitude alteration: $\tilde{\mathrm{LG}}_{0}^{l}(\rho, \phi, w, \tilde{w}) \propto (\rho / \tilde{w})^{|l|}\mathrm{L}^{|l|}_{0}(2 \rho^{2}/\tilde{w}^{2}) \exp{(-\rho^{2}/w^{2})} \exp{(il\phi)}$. The optimal ratio $w/\tilde{w}$ of the Gaussian and polynomial widths for the detection masks have been calculated from the experimentally determined fiber mode width and is found to be $w/\tilde{w}=1.6$. The holograms displayed on the first SLM are not modified since the incident Gaussian pump beam width is significantly larger than the width of the corresponding mask.

\section*{Optimization algorithm}
\begin{figure}
	\center{\includegraphics[width=0.95\linewidth]{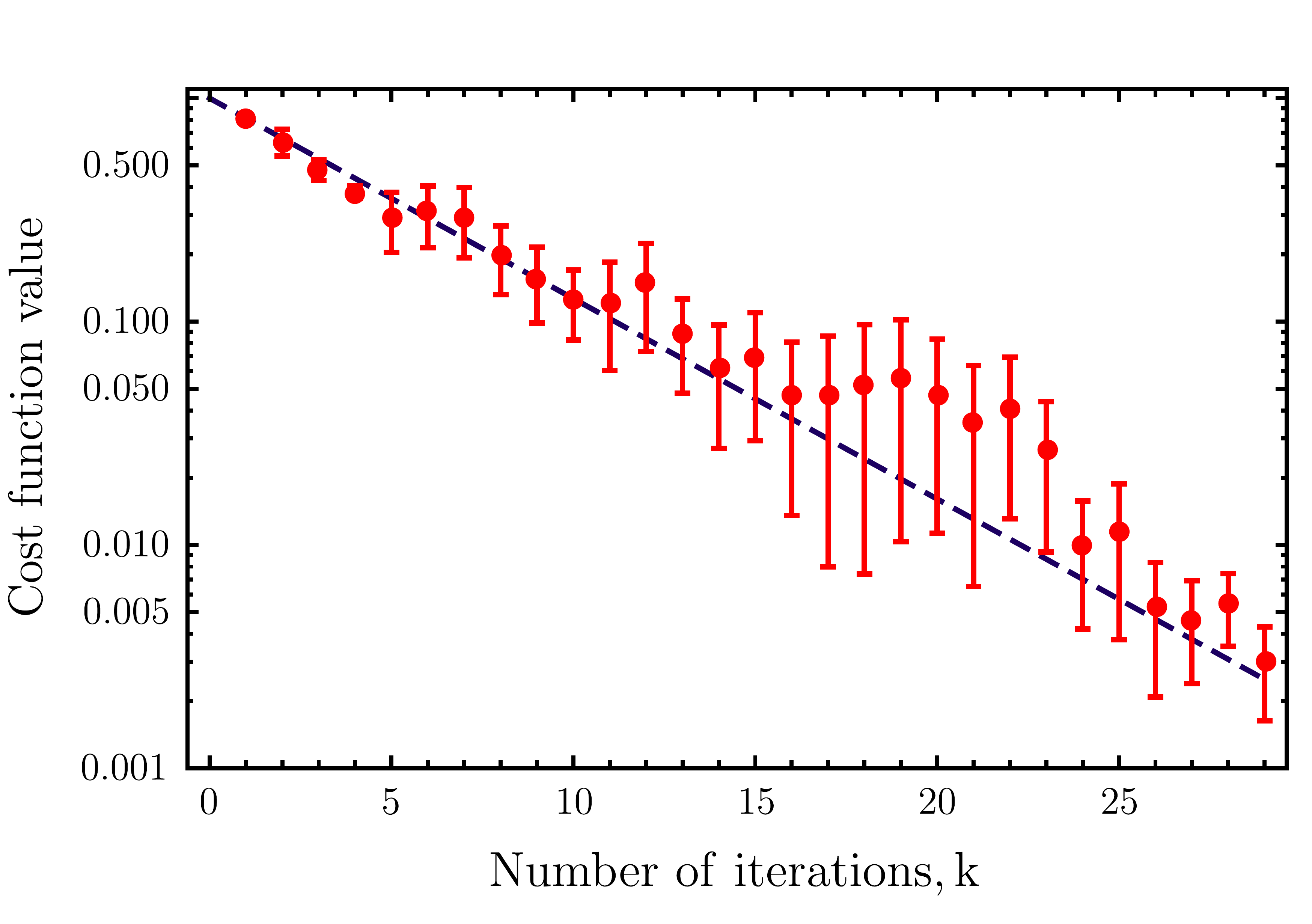}}
	\caption{Measured cost function value as a function of the number of iterations $k$ for the SPSA algorithm. Dots represent the experimental data, the dashed line is a fit with $\exp{(-\gamma k)}$ function with $\gamma=0.20 \pm 0.01$ (dashed line).}\label{fig:spsa}
\end{figure}
As mentioned in the main text, we implement the SPSA algorithm to optimize the spatial mode structure of the pump beam. The dependence  of the cost function value on the number of algorithm iterations for the case of maximally entangled qutrits is depicted in Fig.\ref{fig:spsa}. One may see, that the cost function converges exponentially to $4 \times 10^{-3}$ value in 30 iterations. The presented dependence is averaged over five different runs of the algorithm starting with a Gaussian pump. Since the chosen cost function is indifferent to the phase of the prepared states, the resulting pump beam profiles were found to be similar to each other except for the global rotation about the beam propagation axis. With the doubled number of iterations and increased exposure time, we have achieved the cost function value around $10^{-5}$, which is comparable to the experimental error. This final configuration of the pump beam profile was used in the experiment.

Despite the fact, that SPSA is not a global search algorithm, our experiment does not reveal possible problems associated with trapping in local minima. The optimal values of the algorithm parameters were found to be $a=1$, $c=0.01$, $\alpha=0.6$ and $\gamma=0.1$ (in the notation of \cite{spall1998implementation}). 
\begin{figure}
	\center{\includegraphics[width=0.95\linewidth]{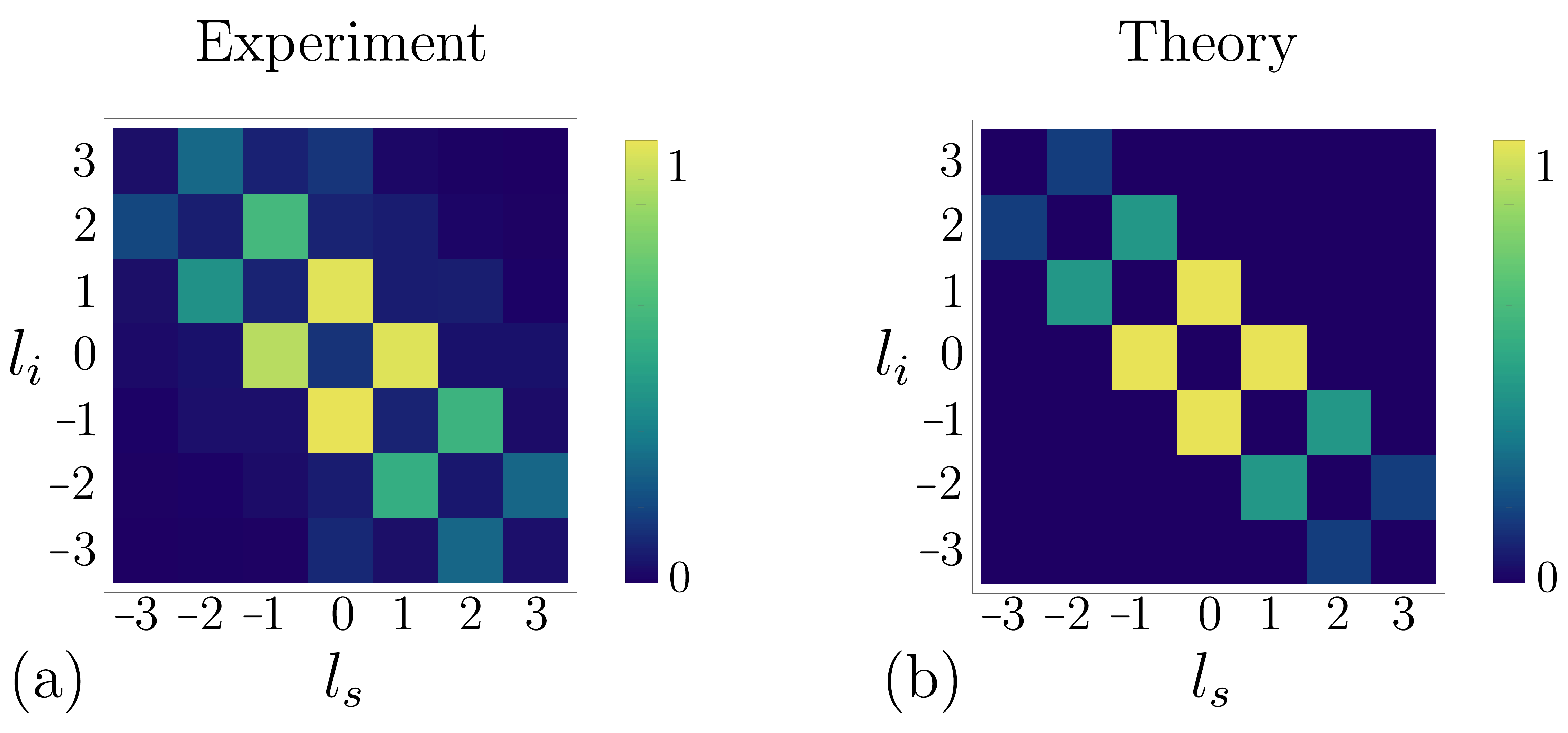}}
	\caption{Spiral spectrum obtained (a) experimentally and (b) from numerical simulations. The four central peaks correspond to the state of two maximally entangled qubits (the spatial Bell state) generated by pumping the crystal with a first order Hermite-Gaussian beam.}\label{fig:bell}
\end{figure} 

\section*{Bell Inequalities for qutrits}
According to \cite{collins2002bell}, the strongest violation of generalized Bell inequalities for a maximally entangled qutrits $\ket{\Psi^{(3)}} = (1/\sqrt{3}) \sum_{l=-1}^{1} \ket{l,l})$ is obtained for the projective measurement on the following states in the signal channel $A$ and the idler channel $B$:
\begin{align}
	&\ket{n}_{A,a}=\dfrac{1}{\sqrt{3}} \sum_{l=-1}^{1} \exp \Big(i \dfrac{2\pi}{3} l (n+\alpha_{a}) \Big) \ket{l}_{A}, \label{MaxVec:1} \\
	&\ket{m}_{B,b}=\dfrac{1}{\sqrt{3}} \sum_{l=-1}^{1} \exp \Big(i \dfrac{2\pi}{3} l (-m+\beta_{b}) \Big) \ket{l}_{B}, \label{MaxVec:2}
\end{align}
where $a,b=1,2$ enumerate the possible choices of measurements, and $\alpha_{1}=0$, $\alpha_{2}=1/2$, $\beta_{1}=1/4$, $\beta_{2}=-1/4$.

Taking into account the non-zero phases $\phi_{1}=\theta_{1}-2\theta$ and $\phi_{2}=\theta_{2}+2\theta$ of corresponding terms $\ket{\scalebox{0.8}[1.0]{-}1,\scalebox{0.75}[1.0]{-}1}$ and $\ket{1,1}$ in (\ref{QutritPhases}), we rewrite (\ref{MaxVec:1}) and (\ref{MaxVec:2}) with the following substitutions: 
\begin{align}
	\ket{-1}_{A,B} &\rightarrow \exp(i\phi_{1}/2)\ket{-1}_{A,B},\nonumber \\
	\ket{1}_{A,B} &\rightarrow \exp{(i\phi_{2}/2)}\ket{1}_{A,B}
\end{align}
with $\phi_{1}$ and $\phi_{2}$ retrieved from the full state tomography. Now, using the probabilities $P(A_{a}=n, B_{b}=m)$, we are able to find Bell parameter
\begin{eqnarray}
\label{BellParameter}
	I_{3} = &+& [P(A_{1}=B_{1})+P(B_{1}=A_{2}+1) \nonumber \\  
	&+&P(A_{2}=B_{2}) +P(A_{2}=B_{1}) ] \nonumber \\ 
	&-&[P(A_{1}=B_{1}-1)+P(B_{1}=A_{2}) \nonumber \\ 
	&+&P(A_{2}=B_{2}-1) +P(B_{2}=A_{1}-1) ].
\end{eqnarray}
In order to estimate these probabilities we measure the coincidence counts rates for 36 different projections $|\bra{n_{A,a},m_{B,b}} \Psi^{(3)} \rangle |^{2}$ with all possible $n,m=0,1,2$. In particular, 24 of these probabilities are used to calculate $I_{3}$ and the additional 12 of them are used to normalize all the outcomes by the total number of coincidences.
\begin{figure}
	\center{\includegraphics[width=0.95\linewidth]{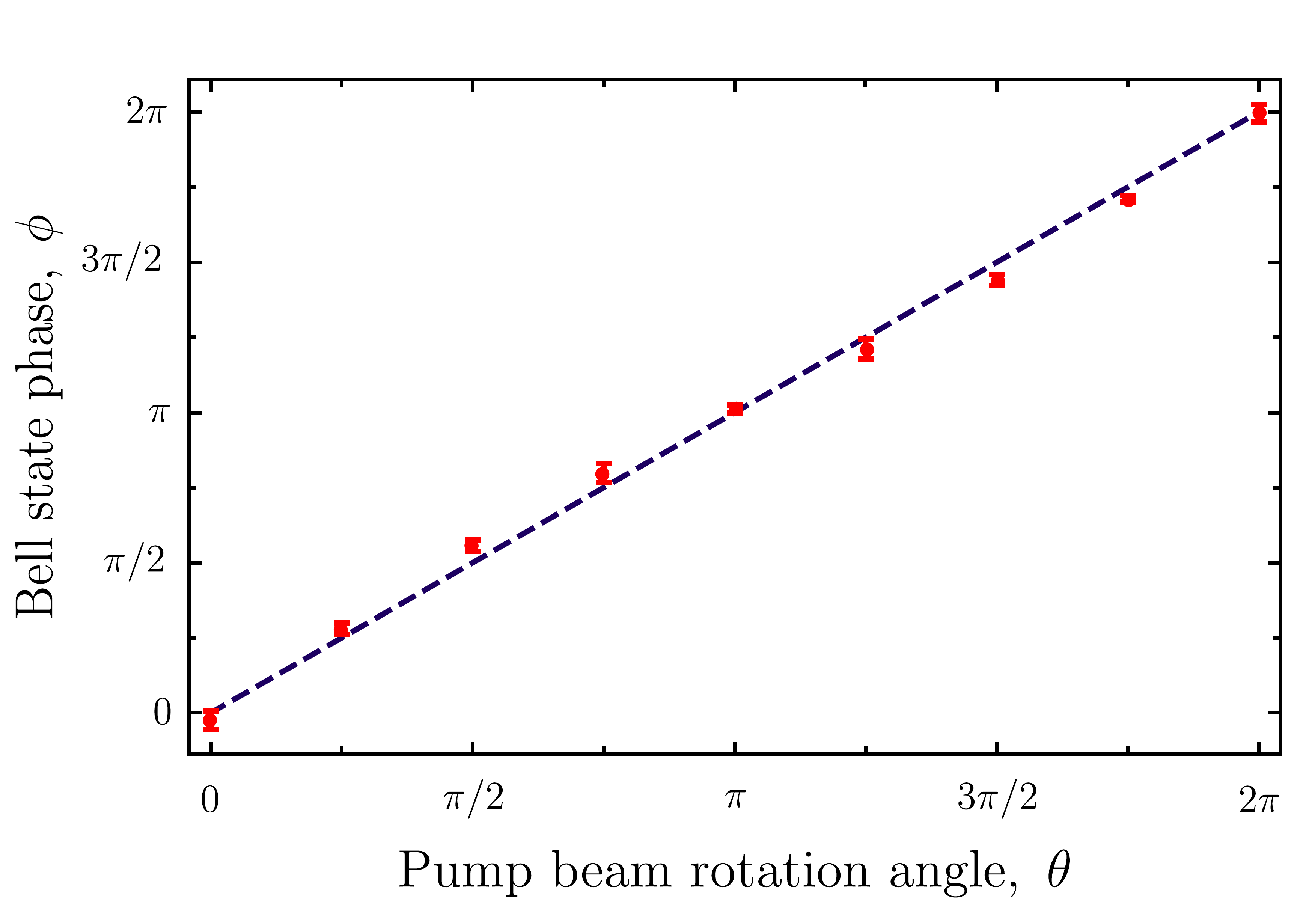}}
	\caption{The dependence of the phase $\phi$ of the spatial Bell state on the pump beam rotation angle $\theta$. Dots correspond to the values estimated from the experimental data and the dashed line corresponds to a direct proportional dependence of $\phi$ upon $\theta$.}\label{fig:bell_phases}
\end{figure}

\section*{Qubits phase control}
It has been shown recently \cite{kovlakov2017spatial} that the use of a first order Hermite-Gaussian beam as a pump leads to the generation of a spatial Bell state of the following form:
\begin{equation}
\label{Bell}
	\ket{\Psi}=\dfrac{\ket{\mathrm{HG}_{00},\mathrm{HG}_{10}}+\ket{\mathrm{HG}_{10},\mathrm{HG}_{00}}}{\sqrt{2}},
\end{equation}
where $\ket{\mathrm{HG}_{jk},\mathrm{HG}_{ut}} = \ket{\mathrm{HG}_{jk}}_{s} \otimes \ket{\mathrm{HG}_{ut}}_{i}$ and $\ket{\mathrm{HG}_{nm}}$ is a single photon state in the $\mathrm{HG}_{nm}$ mode. This Bell state can be easily rewritten in terms of OAM states $\ket{l}$ as
\begin{equation}
\label{Bell_OAM}
	\ket{\Psi}=\dfrac{\ket{0,1}+\ket{0,\scalebox{0.8}[1.0]{-}1}+\ket{1,0}+\ket{\scalebox{0.8}[1.0]{-}1,0}}{2},
\end{equation}
using the relations $\ket{\mathrm{HG}_{10}} = (\ket{\scalebox{0.8}[1.0]{-}1}+\ket{1})/\sqrt{2}$ and $\ket{\mathrm{HG}_{00}} = \ket{0}$. The measured spiral spectrum for $\mathrm{HG}_{10}$ pump beam followed by the numerical results is presented in Fig.~\ref{fig:bell}.

Ideally, the global rotation of the pump beam by an angle $\theta$ does not affect its spiral spectrum. At the same time, in analogy with the qutrits case, it gives the opportunity to control the phase of two qubits, which are maximally entangled in the subspace spanned by $\ket{0,1}+\ket{1,0}$ and $ \ket{0,\scalebox{0.8}[1.0]{-}1}+\ket{\scalebox{0.8}[1.0]{-}1,0}$:
\begin{equation}
\label{Bell_OAM_2}
\ket{\Psi^{(2)}(\phi)}=\dfrac{ \ket{0,1}+\ket{1,0}}{2}+\exp{(i \phi)} \dfrac{\ket{0,\scalebox{0.8}[1.0]{-}1}+  \ket{\scalebox{0.8}[1.0]{-}1,0}}{2}.
\end{equation}
Our experimental results reveal that this phase $\phi$ is directly proportional to the global rotation angle $\theta$ (see Fig.~\ref{fig:bell_phases}). To estimate the phase of the produced state for each rotation angle of the $\mathrm{HG}_{10}$ pump beam we perform state tomography in the same 9-dimensional subspace as for the qutrits. Since the change of qubitы amplitudes during rotation is insignificant, we do not use the adaptive algorithm to equalize them.

	
\begin{figure}
	\center{\includegraphics[width=1.\linewidth]{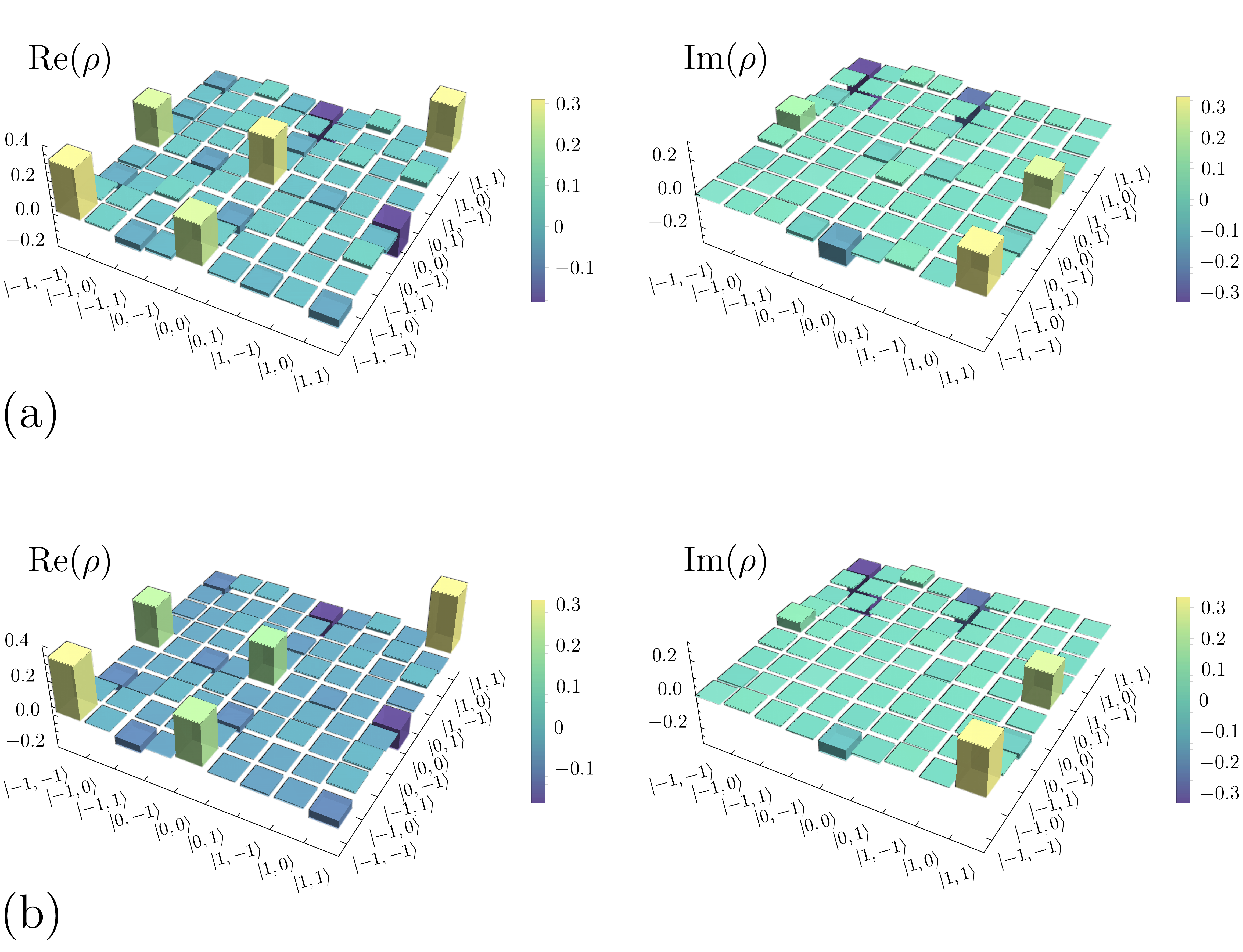}}
	\caption{Density matrices of the reconstructed qutrit states. (a) A maximally entangled qutrit state, (b) a nonmaximally entangled state $\ket{\Psi}=1/\sqrt{2+\gamma^2}\left(e^{i \phi_1} \ket{-1,-1} + \gamma\ket{0,0} + e^{i \phi_2}\ket{1,1}\right)$ with $\gamma=0.79$.}\label{fig:tomogrpahy}
\end{figure}
 
\section*{Full state tomography}

An estimate for the density matrix was obtained by minimizing the $\chi^2$ quantity for the predicted probabilities $p^{(P)}_{i}$ and the measured probabilities $p^{(M)}_{i}\propto R_{c}$ (the coincidence rates $R_{c}$ were normalized to sum to unity):

\begin{equation}
\chi^{2}=\sum^{N}_{i=1} \dfrac{\left(p^{(M)}_{i}-p^{(P)}_{i}\right)^{2}}{p^{(P)}_{i}}.
\end{equation}

For the reconstruction in a $3\times 3$ dimensional subspace, we describe an estimate for a density matrix by a Cholesky decomposition of a $9\times 9$ matrix with 80 independent real parameters. Thus the reconstructed matrix is Hermitian and positive semidefinite with unit trace by construction \cite{KwiatPRA2001}. We also chose an over-complete set of measurements to increase the accuracy of the reconstruction. The state was projected onto a set of $N=15 \times 15$  eigenvectors of the generalized Gell-Mann matrices with the modes $\ket{-1}$, $\ket{0}$ and $\ket{1}$ as a measurement basis. The results of the experimental state reconstruction of a maximally entangled and non-maximally entangled qutrits states are shown in Fig.~\ref{fig:tomogrpahy}. 

\end{document}